\documentstyle[aps,prl,floats,graphicx,epsfig]{revtex}

\begin{document}
\draft

\twocolumn[\hsize\textwidth\columnwidth\hsize\csname @twocolumnfalse\endcsname

\title{ Event horizons and ergoregions in
$^3$He }
\author{ T.A. Jacobson$^{1}$ and G.E. Volovik$^{2,3}$}
\address{
$^{1}$Department of Physics, University of
Maryland\\ College Park, MD 20742-4111, USA\\
$^{2}$Low Temperature Laboratory, Helsinki University of
Technology\\
P.O.Box 2200, FIN-02015 HUT, Finland\\
$^{3}$L.D. Landau Institute for
Theoretical Physics, \\ Kosygin Str. 2, 117940 Moscow, Russia }

\date{\today}
\maketitle

\begin{abstract}
Event horizons  for fermion quasiparticles naturally arise in
moving textures in superconductors and Fermi superfluids. We discuss
the example of a planar soliton moving in superfluid $^3$He-A,
which is closely analogous to a charged rotating black hole.
The moving soliton will radiate quasiparticles via the Hawking effect
at a temperature of about 5 $\mu$K, and via vacuum polarization
induced by the effective `electromagnetic field'
and `ergoregion'.  Superfluid $^3$He-A thus appears to be a useful system
for experimental and theoretical simulations of quantum effects
related to event horizons and ergoregions.\end{abstract}

\


\

] \narrowtext
%\twocolumn

\section{ Introduction}.

A black hole event horizon is the causal boundary of the exterior
region of spacetime.
According to quantum mechanics,
an event horizon emits Hawking
radiation\cite{Hawking} and possesses a Bekenstein
entropy\cite{Bekenstein}. These phenomena lie at the
intersection of gravity and quantum mechanics, and have
played a major role in efforts to understand quantum gravity.
However, the nature of the fundamental degrees of freedom and the
physics at short distances is still not understood,
so basic questions about the statistical
meaning of black hole entropy, the back-reaction to Hawking radiation,
the origin of the outgoing modes, and unitarity, remain
unresolved. Moreover, the effect of the Hawking radiation is
negligible
for the physics of solar mass black holes,
since the temperature of the
Hawking radiation $T_H$ is extremely small ($\sim 10^{-7}$ K).
The only conceivable experimental consequences of the Hawking
radiation at present would arise from evaporation of a (hypothetical)
population of primordial black holes.

For this reason models simulating
event horizons in condensed matter can be useful.
The first attempt at a model of this kind was made with a  moving
liquid \cite{UnruhSonic,Jacobson1991,Visser1997}. The propagation of
sound
waves on the background of a moving inhomogeneous liquid is similar to the
propagation of light in (3+1)-dimensional Lorentzian geometry, and is governed
by the relativistic wave equation:
$$\partial_\mu\left(\sqrt{-g}g^{\mu\nu}\partial_\nu\Psi\right)=0~~.\eqno(1.1)$$
The `acoustic'  metric $g^{\mu\nu}$, in which the sound wave is
propagating, is determined by the inhomogeneity and local flow velocity of the
liquid.  If the liquid moves supersonically a sonic `event horizon' can
arise. A drawback of this model for the simulation of black hole
physics
is that ordinary liquids are essentially dissipative systems and are
very far from the
condition where quantum effects can be of any importance: this smears the
effects that, like the Hawking effect, are related to quantum fluctuations.

Better candidates are superfluids, which allow
nondissipative  motion of the vacuum (superfluid condensate,
or ground state), and which also support well defined elementary excitations
that propagate in a `curved space' of the inhomogeneous moving condensate.
The Fermi superfluids (including superconductors) are appealing
candidates because their
low temperature dynamics are described by quantum field theories similar to
those in high energy physics\cite{Wilczek}.
Among them superfluid $^3$He-A has the
advantage that this superfluid supports an effective gravity caused by some
components of the superfluid order
parameter\cite{Exotic}.

There is one important obstacle to the formation of an horizon in a
moving condensate: superfluidity collapses, i.e. the condensate
disappears, before the corresponding speed of light is reached. For example in
superfluid $^4$He the Landau velocity at which the  condensate is
unstable to roton excitation is smaller than the speed of sound and
thus the supersonic flow can not be established. For
fermionic systems the collapse of  superfluid/superconducting state due to
``superluminal" motion of the condensate was discussed in
\cite{KopninVolovik1998a}.
Therefore we have looked for a model in
which the condensate is at rest with respect to the  container.

We show here that a `superluminally' moving inhomogeneity of the order
parameter (soliton, vortex or other texture) in superfluid $^3$He-A  provides
such a model and can simulate the physics of
an event horizon and ergoregion for `relativistic'
massless fermions---the Bogoliubov-Nambu  quasiparticles. The
`superluminally' moving soliton produces dissipation due to quantum
radiation of the fermions via several mechanisms, which decreases  the
soliton velocity. Similar processes
occur also for a charged, rotating, black hole, where the
Schwinger pair production, pair production in the ergoregion outside
the horizon, and Hawking radiation lead to discharge, spin-down, and
evaporation of the black hole. So both the `superluminally' moving soliton and
the black hole are quasi-equilibrium, unstable inhomogeneous states
exhibiting an event
horizon.

\section{ Relativistic fermions in $^3$He-A.}

The spontaneous breaking of symmetry in the superfluid condensate in $^3$He-A
is characterized in part by a unit vector $\hat l$, which
points along the spontaneous angular momentum of the Cooper pairs and
determines
the direction of the zeroes in the energy spectrum of the
Bogoliubov-Nambu fermion quasiparticles \cite{Vollhardt1990,Exotic}
$$E(\vec p)=\pm\sqrt{ v_F^2(p-p_F)^2 +{\Delta_A^2\over p_F^2}(\hat
l \times\vec p)^2}~~.\eqno(2.1)$$
Here $v_F(p-p_F)$ is the quasiparticle energy in the normal Fermi liquid above
transition, with $p_F$ the Fermi momentum and $v_F=p_F/m^*$; $m^*$ is the
effective mass, which is of order the mass $m_3$ of the $^3$He atom;
$\Delta_A$ is the so-called gap amplitude.

The energy in Eq.(2.1) is zero at two points $\vec p
= e \vec A$ with $\vec A= p_F\hat l$ and $e=\pm 1$. Close to the two zeroes
of the energy spectrum one can expand in $\vec p
- e \vec A$ and the spectrum $E(\vec p)$  becomes
that of a charged, massless relativistic particle propagating
in a curved spacetime in the presence of an electromagnetic vector
potential:
$$ g^{\mu\nu}(p_\mu -eA_\mu) (p_\nu -eA_\nu)=0~~.\eqno(2.2)$$
The `four-momentum' $p_\mu$, `electromagnetic vector potential' $A_\mu$,
and inverse `metric tensor' $g^{\mu\nu}$ in this covariant expression
are specified by giving their components in the coordinate system
$(x^0,x^i)$ where $x^0$ is the Newtonian time $t$ and $x^i=$
are Cartesian spatial coordinates
at rest with respect to the superfluid:
$$(p_0,\; p_i)=(-E,\;  p_i),\eqno(2.3a)$$
$$(A_0,\; A_i)=(0,\;  p_F  l_i),\eqno(2.3b)$$
$$g^{00}=-1,~~~~g^{0i}=0,~~~~
g^{ik}= c_\perp^2 (\delta^{ik} -   l^i  l^k) +  c_\parallel^2   l^i
 l^k ~~.~~ \eqno(2.3c)$$
$E$ and $p_i$ on the right hand side of (2.3a) are the
Newtonian energy and momentum, and the index on $p_i$ and $l_i$ on the
right hand sides of (2.3a) and (2.3b) is lowered with the Euclidian metric
$\delta_{ij}$.  (Once the ``relativistic" quantities are defined by (2.3),
the Euclidean metric plays no further explicit role in the dynamics.)
The fermion quasiparticles actually satisfy the curved
spacetime  Weyl equation for massless charged chiral
spinors\cite{Exotic,Volovik1986},
although for our purposes here all that is needed is the dispersion
relation (2.2). In general there is an additional term in the vector
potential\cite{Volovik1986}
which is proportional to the gradient of the $\hat l$ vector.
Also, the square of the Weyl equation contains extra terms: in symbolic
form it is of the type   $(p-eA)^2 + R + \sigma\cdot F =0$. Here $R$ is the
Ricci scalar, $F$ is the electromagnetic field strength,
and $\sigma$ is the spin (in our case it is the
Bogoliubov-Nambu ``spin" of quasiparticles in particle-hole space).
We ignore all these extra terms in the dispersion relation (2.2),
since they are proportional to the gradients of the $\hat l$ vector
and thus are small in the $\hat l$ texture discussed here:
$F \propto \nabla \hat l$, $R \propto (\nabla \hat
l)^2$. In principle  these terms affect the propagation of the field,
however they play no essential role in the particle production processes
studied here.
Hereafter we omit the quotes when referring to the quasi gravitational and
electromagnetic fields, since no actual such fields  enter our problem.

The quantities $c_\perp= \Delta_A / p_F$  and $c_\parallel=v_F$
in the inverse metric (2.3c) are `speeds of light'
propagating transverse to $\hat l$ and along $\hat l$ correspondingly.
The magnitudes of the $^3$He-A parameters at zero pressure are:
$m^*\simeq 3m_{\rm He-3}$, $\Delta_A\simeq 1.7 {\rm mK}$, $v_F\simeq 55 {\rm
m/s}$, $\Delta_A / p_F \simeq 3 {\rm
cm/s}$. As a result the `speed  of light' is very anisotropic
(in Cartesian coordinates):
$c_\perp\simeq 0.5\cdot10^{-3} c_\parallel $.
The relativistic approximation (2.2) is valid provided $p-p_F\ll p_F$
and $p_\perp=|\vec p\times\hat l| \ll m^*c_\perp$. The condition
$E\ll m^*c_\perp^2\sim 0.5\cdot 10^{-3}T_c \sim  0.5\mu$K
is thus sufficient for this approximation.
This upper limit is still significantly  lower than the lowest
confirmed temperature reached so far in superfluid $^3$He experiments,
about $100\mu$K. The actual lower bound is probably lower than
this but at the moment there is
no reliable thermometry below $100\mu$K \cite{Bunkov}.
If the energy is  higher than $m^*c_\perp^2$, `non-relativistic'
higher order corrections must be added in general.
However there are many examples (such
as axial anomaly and zero charge effect) where only
the propagation along the $\hat l$-axis
is important, in which case the only restriction is that $T\ll
T_c$ so that the thermal fermions are concentrated in the
vicinity of the nodes.

If the $\hat l$ texture moves with constant velocity
$\vec v$, then to obtain manifest time-independence of the
background one must use
the coordinate system which is at rest with respect to the texture.
Let us from now
on  denote the coordinates in the texture frame by the unprimed
letters $(t,x^i)$, and those
in the superfluid frame by the primed letters
$(t',x'^i)=(t,x^i+v^i t)$, where $v^i$ is the velocity of the texture.
The dispersion relation
in the moving frame is obtained from (2.2) and (2.3a,b,c)
simply by finding the
components of the tensors $p_\mu$, $A_\mu$, and $g^{\mu\nu}$ in the
new coordinate system\footnote{We use here the fact that, under the
Galilean transformation of coordinates,
the tensor transformation law for the covariant ({\it not} contravariant)
4-momentum agrees with the Galilean transformation law for the
energy and momentum of quasiparticles, so the resulting components
of $p_\mu'$ are in fact the correct Galilean components. That is,
it is not necessary to transform back to the rest frame of the
superfluid in order to correctly identify the Galilean energy and momentum.}:
$$(p_0,\; p_i)=(-E'+p'_iv^i,\; p'_i),\eqno(2.4a)$$
$$(A_0,\; A_i)=(p_F l_iv^i,\; p_F l_i),\eqno(2.4b)$$
$$g^{00}=-1,~~~~g^{0i}=v^i,~~~~
g^{ik}= c_\perp^2 (\delta^{ik} -   l^i  l^k) +  c_\parallel^2   l^i
 l^k -v^iv^k~~.~~ \eqno(2.4c)$$
Note that $-p_0=E=E'-p'_i v^i$ is just the energy $E$
of the quasiparticle in the moving frame.  In the moving frame the
metric tensor, and electromagnetic vector potential do not depend on time and
thus the quasiparticle energy $E$ is {\em conserved}.
The condition $|E'|\ll m^*c_\perp^2$ which ensures the
validity of the  relativistic approximation
becomes, in terms of $E$, $|E+p_iv^i|\ll m^*c_\perp^2$.

\section{ Moving soliton.}

According to Eq.(2.2) the low-energy fermions move in gravitational and
electromagnetic fields simulated by
the $\hat l$ texture and its   velocity  $\vec v$.
As explained above, to avoid the collapse of the condensate it is important to
have the superfluid at rest with respect to the container, so we do not
introduce
the  velocity field $\vec v_s$. While the superflow is limited by the
critical velocity $\sim c_\perp$
above which the $^3$He-A phase
is destroyed and the broken symmetry is restored, the velocity
$v$ of the texture can  exceed $ c_\perp$ and can approach $v_F$
\cite{LeggettYip,Schopohl}. The $\hat l$-solitons are
well resolved in NMR experiments \cite{VortexSheetExp}, and pulsed NMR
can be used to accelerate them.

Here we consider a topologically stable texture, a ``domain wall" soliton
moving with velocity $v$ in the $z$ direction. We choose the so-called splay
soliton\cite{Vollhardt1990}
$$\hat l= \hat z \cos \alpha(z) +  \hat x \sin \alpha(z)~~,~~z= z'-vt
~~,\eqno(3.1)$$ where $z$ is the coordinate co-moving with the soliton and
$z'$ is the coordinate in the superfluid frame. Since the exact structure of
the realistic soliton\cite{Vollhardt1990}
is not important for our purposes,  we consider a
simplified profile for this soliton,
$$\hat l= -\hat z\;
\tanh{z\over d}  +  \hat x\; \mbox{sech}\,
{z\over d}~~.\eqno(3.2)$$
The thickness $d$ of the
soliton is on the order of the so-called dipole length
$\xi_D\sim 10~\mu$. The profile of the soliton is shown in
Fig. 1.

%%%%%%%%%%%%%%%%%%%%%%%%%%%%%%%%%%%%%%%%%%%%%%%%%%%%%%%%%%%
\begin{figure}[!!!t]
%\centerline{\epsfxsize=0.40\textwidth\epsfbox{SolitonFig1.eps}}
%\bigskip
\begin{center}
\leavevmode
\epsfig{file=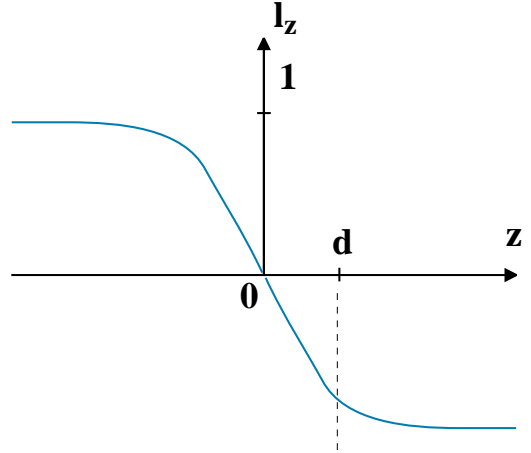,width=0.8\linewidth}
\caption[soliton1]
    {Profile of the $z$-projection of the vector $\vec l$ across the
 splay soliton wall in superfluid $^3$He-A.
  This soliton is topologically stable and can
  propagate in the liquid without destruction.}
\label{soliton1}
\end{center}
\end{figure}
%%%%%%%%%%%%%%%%%%%%%%%%%%%%%%%%%%%%%%%%%%%%%%%%%%%%%%%%%%

The vector potential and metric tensor in the frame comoving with the soliton
do not depend on time, nor do they depend on the $x$ or $y$ coordinates.
Thus the energy $E=-p_0$ and momentum components $p_x$ and $p_y$ of the
fermions
are conserved quantities. The equations (2.4) and (3.1) give the following
nonzero components for the vector potential:
$$A_0=vp_F\cos\alpha,~~~~A_x=p_F\sin\alpha,~~~~A_z=p_F\cos\alpha~,\eqno(3.3)$$
and for the inverse metric:
$$g^{00}=-1,~~~~g^{0z}=v,~~~~g^{yy}=c_\perp^2,
$$
$$g^{zz}=c_\perp^2\sin^2\alpha +v_F^2\cos^2\alpha -v^2~,~ $$
$$g^{xx}= c_\perp^2 \cos^2\alpha +v_F^2
\sin^2\alpha,~~~~g^{zx}=(v_F^2- c_\perp^2) \sin \alpha  \cos \alpha~.~
\eqno(3.4)$$

\section{ Pair production in electromagnetic field.}

The vector potential associated with the moving
soliton gives rise to both magnetic and electric fields. In the
soliton frame the electromagnetic field strength tensor
$F_{\mu\nu}=\partial_\mu A_\nu-\partial_\nu A_\mu$ has the nonzero
components
$$F_{zx}=p_F \partial_z\sin\alpha,
~~~~F_{0z}=vp_F \partial_z\cos\alpha~.\eqno(4.1)$$
The invariant combination
$$``{\bf B}^2-{\bf E}^2~"={1\over
2}F_{\mu\nu}F_{\alpha\beta}g^{\mu\alpha}g^{\nu\beta}= v_F^2c_\perp^2
F_{zx}^2\left(1 -{v^2\over v_F^2\cos^2\alpha}\right)~~\eqno(4.3)$$ does not
depend on the coordinate frame. For any velocity $v$ there are two planes
$z=\pm
z_p$,
$$\cos^2\alpha(z_p)= {v^2\over v_F^2} ~~, \eqno(4.4)$$
where the magnitude of the electric field equals the
magnitude of the magnetic field. In the region between these planes, where
${\bf E}^2>{\bf B}^2$, the electric field induces the Schwinger production of
pairs of fermions \cite{Schwinger}. This leads to dissipation during the motion
of the soliton, which gives rise to a friction force on the soliton even at
zero
temperature and the soliton will decelerate.

For textures where
$A_0(z)$ has equal asymptotes $A_0(\infty)=A_0(-\infty)$ at both
infinities the
situation is different. In this case the  potential $ \Phi(z)=A_0(z)- A_0(\pm
\infty)$ represents a potential well for the fermions. The
fermions formed by Schwinger radiation finally occupy all the negative energy
states in this potential well. After that the radiation stops. The filling of
the negative energy levels will lead to a modification of the vacuum in the
vicinity of the soliton.  After that the soliton with the modified structure
will move without friction. In our case the potential well is unbounded,
$A_0(\infty)=-A_0(-\infty)\neq A_0(-\infty)$. The negative energy levels cannot
be filled, thus the radiation will lead to the deceleration of the soliton
until
it reaches zero velocity.

\section{ Horizon and ergoregion for the fermions.}

If the velocity $v$ of the soliton exceeds $c_\perp$, the metric (3.4)
describes a
planar ``black hole" with an ergoregion outside the horizon, or rather a black
hole/white hole pair. In the frame of the soliton,
the horizons are lightlike surfaces at fixed position $z=\pm z_h$. These
are given by an equation $f(z)=0$, where
the gradient $\partial_\mu f$ is lightlike:
$0=g^{\mu\nu}\partial_\mu f\partial_\nu f=g^{zz}(\partial_z f)^2$.
That is, at the horizons one has $g^{zz}=0$, or
$$c_\perp^2
\sin^2\alpha(z) +v_F^2
\cos^2\alpha(z) =v^2~~. \eqno(5.1)$$
Thus
$$\cos^2\alpha(z_h)=\frac{v^2-c_\perp^2}{v_F^2-c_\perp^2}~.\eqno(5.2)$$

The physical meaning of these horizons is revealed if one
introduces $c^{z}$, the `speed of light in the $z$ direction' in
the superfluid rest frame. Any
planar lightlike surface (``wavefront") is given by an equation $f(t,z')=0$,
where $z'=z+vt$  is the coordinate in the superfluid frame, and
$c^{z}$ is defined by  $(\partial_t  +c^z\partial_{z'}) f=0$.
The lightlike condition on the gradient $\partial_\mu f$
yields $c^{z}=(g'^{z'z'})^{1/2}$. Since
$g'^{z'z'}=g^{zz} + v^2$, the horizons (at $g^{zz}=0$ in Eq.(5.2)) occur where
the speed of light equals the velocity of the soliton, $c^{z}(z_h)=v$. The
speed of particles from the region between the horizons is less than $v$
in the $z$-direction
and thus they cannot propagate out across the leading horizon. Furthermore no
quasiparticle can enter the trailing horizon from the left because as
it approaches its speed drops to the speed of the soliton. So the leading
(future) event horizon  is the black hole and the trailing (past) event
horizon is the white hole (Fig.2).

The horizons appear in the
moving soliton if $v>c_\perp$.
Note that the ${\bf E}^2>{\bf B}^2$ region (4.4)
extends a bit outside the horizon, and exists even when $v<c_\perp$
and there is no horizon.

%%%%%%%%%%%%%%%%%%%%%%%%%%%%%%%%%%%%%%%%%%%%%%%%%%%%%%%%%%%
\begin{figure}[!!!t]
%\centerline{\epsfxsize=0.40\textwidth\epsfbox{SolitonFig2.eps}}
%\bigskip
\begin{center}
\leavevmode
\epsfig{file=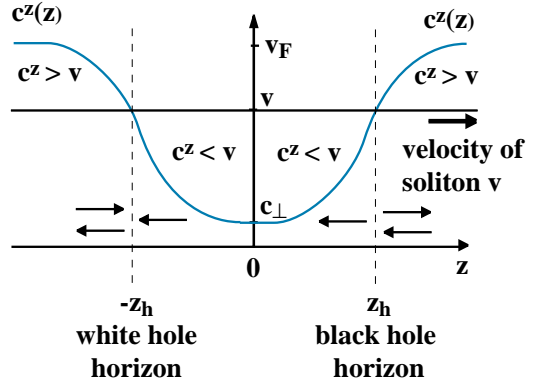,width=0.8\linewidth}
\caption[soliton2]
    {The ``speed of light" in the $z$-direction in the superfluid frame,
$c^z(z)$.
 Quasiparticles in the region $-z_h<z<z_h$, where the speed of light is less
 than the velocity $v$ of the soliton, cannot propagate to the right in the
 frame of the moving texture (the arrows show
 the possible directions of quasiparticle motion in the $z$-direction).
 This region is bounded by the black and white hole horizons. }
\label{soliton2}
\end{center}
\end{figure}
%%%%%%%%%%%%%%%%%%%%%%%%%%%%%%%%%%%%%%%%%%%%%%%%%%%%%%%%%%

The ``ergoregion" is the region where a particle must go faster than
light in order to remain at the same value of $(x,y,z)$. This occurs
where $g_{tt}>0$. In the ergoregion the time translation Killing field
$\partial_t$ is {\em spacelike}, so the conserved ``energy" can be
negative even for a future pointing timelike
four-momentum. As a result the vacuum in the ergoregion is unstable
to creation of pairs of particles, both with future pointing momenta,
with total energy zero. Put differently, the conserved ``energy"  can
be {\it positive} even for a {\it past} pointing
timelike four-momentum. For $^3$He this means that a state in the
occupied valence band (i.e. with the negative sign of the square root in
Eq.(2.1)) has positive energy and thus can tunnel out away from the
ergoregion, leaving behind a negative energy hole state.

The ergoplanes---boundaries of the ergoregion---occur at
$z=\pm z_e$ where $g_{tt}=0$,
which yields
$$\cos^2\alpha(z_e)=
\frac{1-c_\perp^2/v^2}{1-c_\perp^2/v_F^2}~.\eqno(5.3)$$
An ergoregion exists for this soliton only if $v>c_\perp$,
so there is an ergoregion if and only if there is an event horizon.
The ergoplanes lie outside the event horizons (5.2) and outside
the Schwinger pair region (4.4) unless $v$ is extremely close to $c_\perp$
(i.e. unless $v<c_\perp(1-c_\perp^2/v_F^2)^{-1/2}$).

The locations of the boundary of the Schwinger pair production region $z_p$,
event horizon $z_h$, and ergoplane $z_e$ depend on the velocity $v$
of the soliton. To get an idea of these locations and their scale
we have plotted in Fig. 3 the coordinate $z$ versus
$\log(v/c_\perp)$  for each of these three positions.
The Fermi velocity is at the abscissa $\sim 3$
($v_F/c_\perp\sim 10^3$).
Recall that $d\sim 10^5 \AA$, so Fig. 3 shows that
$z_h$ is smaller than $10^3 \AA$ until
$v\sim 10 c_\perp$.

%%%%%%%%%%%%%%%%%%%%%%%%%%%%%%%%%%%%%%%%%%%%%%%%%%%%%%%%%%%
\begin{figure}[!!!t]
%\centerline{\epsfxsize=0.40\textwidth\epsfbox{SolitonFig3.eps}}
%\bigskip
\begin{center}
\leavevmode
\epsfig{file=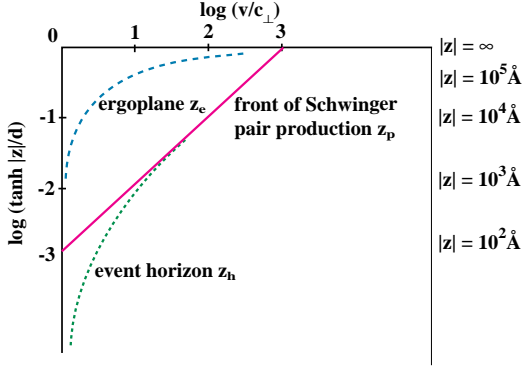,width=0.8\linewidth}
\caption[soliton3]
    {Locations of the horizons, the ergoplanes and the boundaries of the region
  of pair production by the electric field, as a function of the velocity $v$
  of the soliton.}
\label{soliton3}
\end{center}
\end{figure}
%%%%%%%%%%%%%%%%%%%%%%%%%%%%%%%%%%%%%%%%%%%%%%%%%%%%%%%%%%

The horizon has a `transverse velocity' because
the light rays on the horizon are actually moving in the $x$ direction.
This is because the $\hat l$ vector has an $x$-component, so the
speed of light is faster in the $x$-direction on the
horizon. This is analogous to the rotational
velocity of the horizon of a rotating black hole.
To compute this velocity it is helpful to introduce the
``horizon generating Killing field" $\chi^\mu$, which is tangent to the
lightlike curves that generate the horizon.
Since it is spacelike on the horizon, the vector field $\partial_t$
is clearly {\it not} the horizon generating Killing field.
Rather, we have
$$
\chi=\partial_t + w\partial_x ~~,
\eqno(5.4)
$$
where $w$ is some constant which we call the {\it transverse
velocity} of the horizon. Since the horizon is a lightlike surface,
$\chi$ must be orthogonal to $\partial_x$, so
$0=g_{\mu\nu}\chi^\mu(\partial_x)^\nu=g_{tx}+wg_{xx}$.
Thus $w=-(g_{tx}/g_{xx})_h=(g^{zx})_h/v$, or
$$
w=v_F\sqrt{\frac{1-v^2/v_F^2}{1-c_\perp^2/ v^2}}~~.
\eqno(5.5)
$$
A lightlike surface generated by a Killing field is called a
{\it Killing horizon}, so our black hole
horizon is a Killing horizon.
The surface gravity $\kappa$ of a Killing horizon can be
defined by the equation $\partial_a(\chi^2)=-2\kappa\chi_a$,
evaluated on the horizon. Direct but tedious
computation yields
$$
\kappa=\frac{dg^{zz}/dz}{2v}\vert_h =(dc^z/dz)\vert_h
~~,\eqno(5.6)
$$
or
$$
\kappa=\frac{v_F}{d}(1-\frac{v^2}{v_F^2})
\sqrt{\frac{1-c_\perp^2/v^2}{1-c_\perp^2/ v_F^2}}~~.\eqno(5.7)
$$

\section{ Hawking radiation.}

The Hawking temperature
is determined by the surface gravity as
$$T_H={\hbar\over 2\pi k_B} \kappa ~~,\eqno(6.1)$$
where $\kappa$ is given by Eq.(5.6) or (5.7).   Note that, as in
the case of Unruh's sonic black hole model, the Hawking temperature
is given by the gradient of a velocity at the horizon. However, in the sonic
case it was the velocity of the fluid, whereas in the present case it is
the (anisotropic) velocity of the fermion quasiparticles.
As long as the soliton velocity $v$ is not too close to either $v_F$
or $c_\perp$,  then $\kappa \simeq v_F/d$, which gives
$T_H\simeq 5\mu$K. This is  an order of magnitude lower than
the lowest confirmed temperature reached in
superfluid $^3$He experiments today, but is
an order of magnitude higher than the temperature
$\simeq 0.5\mu$K above which the nonrelativistic corrections become
important.

The Hawking flux for fermions has the form\cite{Hawking}
$$\Gamma\, \Bigl[\exp\Bigl((E-\mu)/k_BT_H\Bigr)+1\Bigr]^{-1}
\eqno(6.2),$$
where $\Gamma$ and $\mu$
are the emission coefficient and ``chemical potential" for the mode
in question.
In our case, $\mu$ is given by
$$\mu= p_x w  + e A_0(z_h)\eqno(6.3)$$
(neglecting the spin energy), where $w$ is the transverse velocity
of the horizon (5.5) and $A_0(z_h)$ is the scalar potential (3.3)
evaluated at the event horizon.
By way of analogy, for a rotating charged black
hole one has
$\mu= J \Omega + e\Phi$ where $J$ and $e$ are the the angular momentum
and charge of the mode, and
$\Omega$ and $\Phi$ are the angular velocity and the
electric potential of the horizon.
The quantity $E$ in (6.2) is $-p_0$, the conserved energy in
the comoving frame of the soliton, which according to (2.4a)
is equal to $E'-p'_z v$ where $E'$ and $p'_z(=p_z)$ are the energy
and momentum in the frame of the superfluid.
We remind the reader that there is a constraint
$\vert E'\vert \ll m^*c_\perp^2$ on the quasiparticle energy
in order for the relativistic description to be
generally valid. In terms of $E$ this constraint becomes
$|E+p_z v|\ll m^*c_\perp^2$. Since $p_z$ is not conserved
this condition may be satisfied at one point of a quasiparticle
trajectory and not at another. A complete analysis of the
particle creation processes will therefore require the nonrelativistic
treatment in general, although the extent of the nonrelativisitic
corrections will depend on the type of texture and other parameters
of the system.

The existence of Hawking radiation in the black hole case follows from
the assumption that near the horizon the high frequency
outgoing modes of the quantum field  are in the
ground state as defined in a frame falling freely across the horizon.
When the temperature of the heat bath  (normal component of the liquid) is very
low, this assumption holds  for the moving soliton in $^3$He, since the
``freely
falling frame" is the  frame of the superfluid which is at rest with respect to
the container (and thus to the heat bath). The passage of the moving texture
through this frame is essentially adiabatic for the high frequency modes.
As a
result, the distribution of the fermions in the soliton frame
remains thermal and is given by the
fermi-function
$f(E')$ with $E'=E-\vec p\cdot\vec v_n$, where $\vec v_n$ is the velocity
of the
heat bath; and $E$ and $\vec{p}$ are in the soliton frame.

The vacuum is therefore not excited directly by any time dependent forcing,
but it is unstable to tunneling processes arising from both `level crossing'
and the Hawking effect. The level crossing leads to Schwinger pair production
in the `electric' field, as well as pair production in the ergoregion
that would occur even in the absence of electric charge in analogy with
the process outside a rotating black hole. What happens is that the
Fermi sea is `tilted' in space, and some states under the Fermi surface
near the soliton have positive energy relative to the Fermi surface
far from the soliton. Quasiparticles in these states may tunnel out
leaving behind quasiholes (or {\it vice versa}) that are swept past the
horizon.

It was realized\cite{levelxing} shortly after Hawking's discovery
that the flux from pair creation due to level crossing outside the horizon
corresponds to the contribution from states with $E<\mu$ in the
flux formula (6.2). As $T_H\rightarrow0$, the flux is
extinguished for states with $E>\mu$, whereas for states with $E<\mu$
it approaches the nonzero value $\Gamma$, the tunneling probability.
At finite Hawking temperature the flux is modified as indicated by (6.2).

To determine the actual magnitude of the Hawking flux
and ``level crossing flux"
it is necessary to
evaluate the  emission
coefficients (or so-called ``grey-body factors")
$\Gamma(E,p_x,\sigma,e)$ ($\sigma$ is the spin). These
indicate the fraction of each mode that is ``transmitted" from
its high frequency form near the horizon out to infinity, while
the rest is scattered back across the horizon. These coefficients
have not yet been calculated.

\section{ Quantum mechanics
of quasiparticles near the horizon.}

The temperature and chemical potential of the Hawking radiation
were inferred above by exploiting the analogy with Hawking's calculation.
It may be helpful here to exhibit the essential physics
in a simple way\cite{DamourRuffini}.
Neglecting the spin degrees of freedom, the wave equation
for the fermions is the same as in the bosonic case, which
(neglecting the electromagnetic field) is governed
by  the wave equation (1.1).
The outgoing waves oscillate rapidly near the horizon. If we choose
eigenmodes
$$\Psi =\Psi_{E,p_x}(z) e^{-iEt}e^{ ip_x x}e^{ip_y y}~~,\eqno(7.1)$$
and neglect all terms without at least one $z$-derivative, eqn. (1.1)
becomes
$$  [2i(-vE + p_x g^{xz}(z_h))\partial_z+\partial_z(g^{zz}(z)
\partial_z )]\Psi_{E,p_x}=0~~. \eqno(7.2)$$
The general outgoing  solution is:
$$\Psi_{\tilde E}(z) = a \exp\Bigl(2iv\tilde E\int^z dz'/ g^{zz}(z')\Bigr)
$$
$$\simeq a  \exp\Bigl(i(\tilde E/\kappa)\ln(z-z_h)\Bigr)~,\eqno(7.3) $$
with $\tilde E=E-p_xg^{xz}/v=E-p_x w$, where $w$ is the translational velocity
of the horizon (5.4) and $\kappa$ is the surface gravity (5.6).

The outgoing modes that have purely negative frequency with respect to
the free-fall frame (i.e. the superfluid frame) are states below the
Fermi sea, which is in (or near) the
quantum ground state, as discussed above.
Some of these modes have {\em positive} energy in the (stationary) soliton
frame however, so they may tunnel out away from the soliton.
These modes fall into two classes according to whether $\tilde E$ is
less than or greater than zero.  When $\tilde E<0$, the positive energy
states below
the Fermi sea can be located {\em outside} the horizon. Tunneling of these
states is identified as due to level crossing in the ergoregion, and includes
the Schwinger pairs when an electric field is included. When $\tilde E>0$,
the positive energy states below the Fermi sea can only exist {\em behind}
the horizon,
so it might seem that they could never tunnel out. However, this is not
true because these states always have an exponential tail that spills
out across the horizon. That they must have such a tail follows from the
fact that a purely negative frequency wavepacket must be analytic in the
upper half complex time plane. Equivalently, the mode function
$\Psi_{\tilde E}(z)$ must be analytic in the lower half complex $z$ plane.
Analytic continuation of (7.3) across the horizon
in the lower half complex $z$ plane yields
$$\theta(-u)\, \Psi_{\tilde E}(z_h-u)+
e^{-\pi\tilde E/\kappa}\, \theta(u)\, \Psi_{\tilde E}(z_h+u)~,\eqno(7.4)$$
where $u=z-z_h$.
The second term is the tail term. The associated probability current
is the ratio of the squared norm of this piece to the total squared
norm, i.e., it is $(e^{2\pi\tilde E/\kappa}\pm 1)^{-1}$, where
$\pm=+$ for fermions and $-$ for bosons. (For bosons, the term
inside the horizon has negative norm in the relevant inner product.)
The distribution of particles that tunnel across the horizon (in this
sense) is thus a thermal one at the Hawking temperature (restoring $\hbar$)
$T_H=\hbar\kappa/2\pi k_B$
and with the chemical potential
$p_x w$, in agreement with eqns. (6.1) and (6.3).
After tunneling across the horizon (so to speak), the particles are
partially scattered back across the horizon. The fraction that
propagate out to the asymptotic region is the emission coefficient
$\Gamma$ of eqn. (6.2).

\section{ Discussion.}

With this moving texture model it should now be possible to study
some of the questions presented by black hole horizons. The
Hawking radiation, pair production in the ergoregion, and Schwinger
pair production are driven by the kinetic energy of the moving soliton,
and the back reaction will
be to slow the soliton. The Hawking temperature is fairly constant
until, as $v$  approaches $c_\perp$, $T_H$ goes to zero. This is
unlike the evaporation of a neutral black hole which gets hotter
as it shrinks. It is rather like the Hawking radiation from a
black hole with a large magnetic charge which cannot be discharged
and so cools as it evaporates and approaches an extremal black hole.

The radiation from level crossing in the ergoregion
may be observable with current technology, and
the Hawking flux at $\sim$ 5 $\mu$K is probably not too low to be
observed eventually.
The Hawking temperature can be significantly
higher if instead of the soliton one takes a moving planar
interface between $^3$He-A and $^3$He-B. The A-B interface has many
advantages:
it can be moved with high velocity especially at low $T$
\cite{Boyd} and the thickness $d$ of the interface is much shorter $d\sim
500 \AA$. This essentially increases the corresponding ``surface gravity"
and the
Hawking temperature.  But the ``nonrelativistic" corrections
also become more important and this requires further investigations.
On the other hand, for understanding some of the principal issues related to
event horizons, it is not necessary to consider a real system. Gedanken
experiments can be made on model $^3$He-A-like systems, in which the
``nonrelativistic" corrections can be made arbitrarily small.

Horizons can occur in other moving
topological and nontopological textures in
superfluids and superconductors, and for the bosonic degrees of freedom as well
as for the fermions. In particular, the orbital
waves in $^3$He-A  --- oscillations of the $\hat l$ vector --- are
analogous to
electromagnetic waves. At low $T$ their dynamics becomes
relativistic \cite{Exotic} and one can discuss the propagation of such
relativistic bosons in textures with event horizons.

As for other topological objects, an interesting analogy
occurs in the case of quantized vortices, which correspond at
large distances to spinning cosmic strings\cite{Volovik1997}.
The vortex has fermion zero modes bound to the vortex core.
There is a connection between the statistics of these fermion zero modes  and
the fermionic zero modes on fundamental strings, which simulate the
thermodynamics of extreme black holes
\cite{KopninVolovik1998}.  The Hawking radiation is absent if the vortex is
stationary with respect to the heat bath: a stationary
vortex corresponds to a local minimum of the energy and thus no radiation is
possible from this state. If a
vortex moves with respect to the heat bath or if a nonaxisymmetric vortex core
rotates with respect to the heat bath, the spectral flow
of the fermion zero modes lead to dissipation of the vortex
motion and to an additional transverse force on the moving vortex
\cite{KopninVolovik,Stone,KopninVolovik1997}. In some cases this corresponds to
the appearance of a horizon with nonzero surface gravity
\cite{KopninVolovik1998,KopninVolovik1997}.

Another interesting texture is a domain wall in a thin film of $^3$He-A,
where the $\hat l$-vector which is perpendicular to the film
changes sign\cite{SalomaaVolovik1989}.
If this texture is moving the fermion quasiparticles see an effective
2+1 dimensional spacetime with black hole and white hole horizons
and a curvature singularity in between\cite{JacobsonVolovik}. This is in many
ways a much simpler system than the one discussed in this paper, since
there is no ergoregion or (pseudo-)electromagnetic field.

To have a horizon
in a condensed matter system, it is not necessary to create curvature
singularities as inside black holes, since the metric describing the horizon
does not follow from the Einstein equations. Moreover, if curvature
singularities do occur, as in the thin film domain wall texture just mentioned,
the physics is still under control.
The quantum fermions propagating in
the texture obey relativistic dynamics in the low-energy limit and thus
fully exhibit the quantum physics of the horizon, including both the
Hawking radiation and the entropy of the fermion zero modes.
For high enough energies, or inevitably near singularities in the
texture, the fundamental nonrelativistic description takes over.
Thus superfluid
$^3$He is a promising model for experimental and theoretical simulations of
quantum effects related to the event horizon, and may offer useful ideas
about resolving the physics near a singularity.

\section{ Acknowledgements.}

This collaboration was carried out under the EU Human
Capital and Mobility Programme (contract CHGE-CT94-0069). The work of
TJ was also supported in part by NSF grant PHY94-13253, the Institute
for Theoretical Physics at the University of Utrecht, the General Research
Board of the University of Maryland, and the Erwin Schr\"odinger Institute.
The work of GV was also supported in part by
the Russian Foundation for Fundamental Research grant No. 96-02-16072, by the
RAS program ``Statistical Physics'', by the Intas grant 96-0610 and by European
Science Foundation.

 \vfill\eject

\end{document}